# Electroluminescence and photoluminescence of conjugated polymer films prepared by plasma enhanced chemical vapor deposition of naphthalene

M. Rajabi[1,2]*, A.R. Ghassami[1], M. Abbasi Firouzjah[1], S.I. Hosseini[1], B. Shokri[1,2]*

1 Laser and Plasma Research Institute, Shahid Beheshti University, G.C., Evin, Tehran, Iran.

2 Physics Department, Shahid Beheshti University, G.C., Evin, Tehran, Iran.

Thin film Plasma polymerized naphthalene was formed by plasma enhanced chemical vapor deposition as emissive layer. The structure analysis revealed forming a conjugated polymer with a 3-D cross-linked structure. By increasing the plasma power, the optical properties including UV-Vis, photoluminescence and electroluminescence spectra showed two domains with different behaviors.

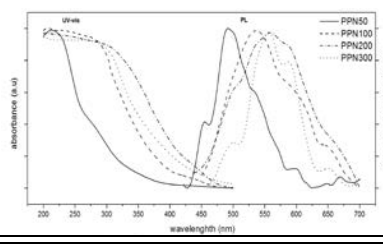

**ABSTRACT :** Polymer light-emitting devices were fabricated utilizing plasma polymerized thin films as emissive layers. These conjugated polymer films were prepared by RF Plasma Enhanced Chemical Vapor Deposition (PECVD) using naphthalene as monomer. The effect of plasma parameters on the structure and optical properties of the conjugated polymers was investigated by applying different plasma powers. The fabricated devices with structure of ITO/PEDOT:PSS/ plasma polymerized Naphthalene/Alq3/Al showed broadband Electroluminescence (EL) emission peaks with center at 535-550 nm. Fourier transform infrared (FTIR) and Raman spectroscopies confirmed that a conjugated polymer film with a 3-D cross-linked network was developed by C-H bond rupture and ring opening process during the plasma polymerization. By increasing the power, ring opening process and cross-linking increased and the products were formed as highly cross-linked polymer films. In addition, as the plasma power increased, the optical properties showed two domains, up to 200 w, the electroluminescence, photoluminescence (PL) and UV-Vis spectra red-shifted and broadened due to increasing the conjugation length, forming more complex polymer and rising the excimeric emissions. At higher powers, a reverse behavior was observed. The conjugation length reduced and a change in the excimeric emission dominance has happened. Also, the relation between the film structure and plasma species was investigated using Optical Emission Spectroscopy (OES).

## Introduction

Conjugated polymers represent a novel class of semiconductors indicating unique optical and electrical properties because of delocalized π band presence in their structure.[1-3] Great potential of forming as flexible and low cost thin films has made them a good option for optoelectronic device applications such as organic light-emitting devices (OLEDs), organic thin film transistors and photovoltaic cells.[4-6] Conjugated polymers combine the optical and the electronic properties of semiconductors with the processing advantages and mechanical properties of polymers. A major advantage of organic semiconductors is that their mechanical and optoelectronic properties can be modified by changing their structure using various polymerization methods and wide range of available monomers.[3,7].

Conventional thin film fabrication methods such as spin coating and dip coating are easy and useful; however, it is difficult to avoid impurity and pin-hole defects in them. These defects have destructive effects on the efficiency of optoelectronic devices.[8,9] On the other hand, chemical vapor deposition (CVD) and Plasma Enhanced CVD (PECVD) are solvent-free processes to form polymer directly from starting monomers.[8] Among these, PECVD is a unique technique for fabrication of high quality and chemically stable thin films.[8,10] Plasma polymerization is initiated by collision between electrically accelerated electrons and monomers. Collisions activate monomers in both gas phase and plasma-surface boundary. Then, activated monomers react with each other and substrate molecules and form a thin layer.[11,12] Thin layers obtained by PECVD are cross-linked, dense, pin-hole free, adherent to various substrates and with less roughness.[10-13] Also, by plasma polymerization, it can be possible to produce various polymers with different structures and thus different mechanical and electro-optical properties from the same monomers. For example, by changing plasma parameters such as pressure and power, the resultant polymer cross-linking density, band gap and luminescent characteristics can be tuned.[12,14,15]

In the past, some few works have been done about plasma polymerization of naphthalene using RF frequency. One more recent work is C.Chang et.al [9] study on RF plasma polymerization of 1-ethylnaphthalene.

In the present study, we prepared conjugated polymer thin films by RF plasma polymerization of naphthalene as the monomer. A vaporizer was used to inject naphthalene vapor to chamber under temperature control to stabilize monomer flow rate and working pressure. We also investigated the effect of plasma power on the polymer structure and optical properties. The FTIR and PL spectra of the resultant polymers were compared. In addition, we reported the fabrication of a multilayer OLED with device structure of ITO/PEDOT:PSS/Plasma Polymerized Naphthalene (PPN)/Alq3/Al. Finally, the EL spectra of different plasma polymerized films were compared for first time.

## Experimental Section

Naphthalene powder 99% from Merck Co. Ltd. was used as the starting monomer without further purification. Indium tin oxide (ITO) coated glass with a resistance of 25/□ from Aldrich Co. Ltd., and quartz sheets were used as the substrate. Substrates were cleaned by ultrasonic using acetone and etha-nol and placed on the ground electrode. First the reaction chamber was evacuated less than 10 mtorr, then naphthalene monomer vapor was injected to the chamber using a vaporizer (Figure 1). The vaporizer stabilized the monomer evaporation rate using a temperature control system. Therefore, working pressure was fixed at 40 mtorr during the polymerization. The glow discharge was generated by a 34 MHz RF generator with a capacitive coupled mechanism and the plasma power was set from 50 to 300 w to form different polymers. Different polymerization times were set for various applying power to obtain layers with thickness of about 150 nm.

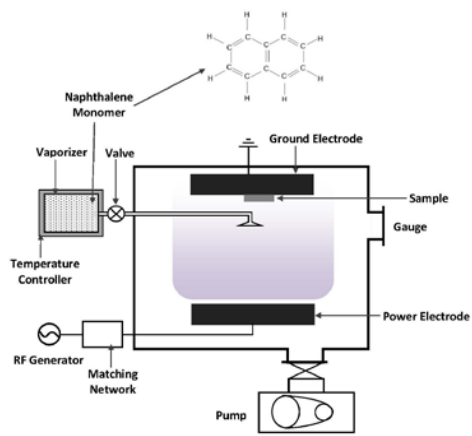

Figure 1. the reaction chamber apparatus

Optical emission spectroscopy (OES) of the Naphthalene plasma was used to understand the plasma environment utilizing an Avantes spectrometer. The chemical structure of thin film polymers was analyzed by FTIR and Raman spectroscopy. The FTIR spectra were taken using Bruker Tensor 27 spectrometer via KBr disc method. Also, the Raman spectra were recorded by an Almega Thermo Nicolet Dispersive Raman Spectrometer with a 532 nm laser beam and 30 mw laser power with 4cm-1 resolution detector. A Shimadzu UV-Vis spectrometer was used to detect UV-Vis absorption spectra. The PL emission spectra were observed by an Avantes spectrometer using a 405 nm laser beam.

For device application purpose, a multi layer OLED was fabricated with following structure: ITO/PEDOT:PSS/PPN/Alq3/Al. Before plasma polymerization, ITO substrates were spin-coated by a PEDOT:PSS thin layer with thickness of about 40 nm. After plasma polymerization of naphthalene, a thin layer of Alq3 with thickness of about 20 nm was coated by thermal evaporation. Finally, an AL layer, as the cathode, with about 200 nm thickness was coated by thermal evaporation. The EL spectra of the devices were measured by using a USB 2000 spectrometer and a Keithley pico-ampermeter.

### Results

In plasma, monomers are activated via energetic electrons. The activation process of naphthalene monomers is initiated with C-H bond rupture reactions. It is expected that by increasing the power, electrons gain more energy. Therefore, bond rupture process was followed by the ring opening of the naphthalene and formed a 3D cross-linked polymer. By further increasing the power, C-H bond rupture and ring opening increased and high energy electrons bombardment caused fracture of double carbon bonds and forming as saturated sp3 bonds. Therefore, polymer chains started to be formed as the highly cross-linked polymer.[9]

Optical emission spectroscopy (OES) was applied during the plasma polymerization for more understanding about the plasma environment. Figure 2 shows the OES spectra corresponding to different plasma powers.

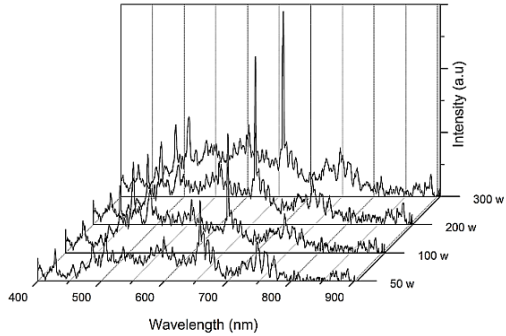

Figure 2. OES spectra of the Naphthalene plasma discharge in different plasma powers

Analyzing the spectra showed the presence of H (Hα at 656.2 and Hβ at 486 nm), C (426.9 and 723.8 nm), CH (431.3 nm) and $H_2$ (506.7 nm) spices in the plasma.[16-19] As the plasma power increased the relative intensity of the most spectral lines increased. It reveals the strong dependence of the spices concentration to the plasma power.[20] Figure 3 shows the relative concentration of some important spices as plasma power increases. Here, each spectrum has been normalized to the 427 nm C peak intensity. By increasing the plasma power, concentration of H and C spices raised due to increasing the fragmentation. It seems the fragmentation caused increasing C-H concentration at low powers, however, by increasing the power and dominance of dissociation process this amount decreased. Increasing C-H dissociation and producing free C spices increase the probability of impacting carbon atoms to each other in the plasma phase and on the film surface which conceive polymerization process.

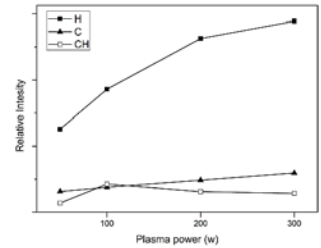

Figure 3. relative intensity of some OES peaks at different plasma power

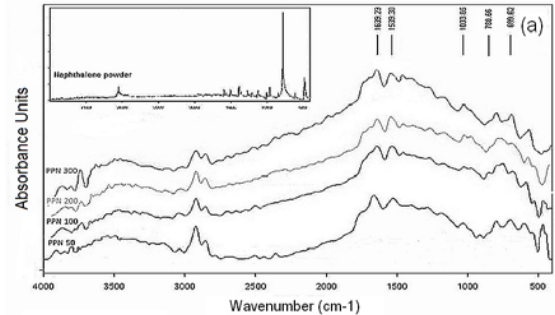

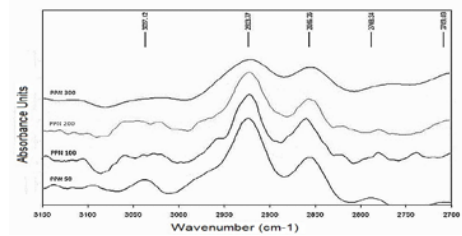

Figure 4) FTIR spectra of the naphthalene monomer and the PPN films prepared in different powers.

Figure 4 illustrates FTIR spectra of the PPN films at different powers. All FTIR spectra showed bands at 780 and 3050 cm-1 that are related to out-of-plane aromatic C-H bending vibration and aromatic C-H stretching vibration, respectively.[6,9,21-24] The peaks in 1500-1600 cm-1 region corresponds to C=C stretching bonds.[9] Also, aliphatic C-H stretching peaks at 2924 cm-1 due to methylene and 2856 cm-1 due to methyl groups and aromatic C-H stretching vibration at 3050 cm-1 were detected in plasma polymerized films.[21,24] Observation of saturated and unsaturated carbons and aromatic struc-ture in FTIR spectra verified the presence of conjugation structure in the plasma polymers. Comparison between the spectra of the PPN films and naphthalene monomer showed reduction of the aromatic C-H bonds during polymerization process. It suggests that the aromatic rings of naphthalene have broken during the plasma polymerization.[24] Also, the presence of aliphatic C-H bonds in polymer films confirms rings opening of naphthalene units and then re-bonding as methyl and methylene groups.[6,9,22,24] By increasing the power, aromatic C-H peaks became weaker due to more ring opening.[24] At high powers, the aliphatic C-H bonds became wider. It seems the C-H bonds have been broken by high energy electron bombardment and residual carbon atoms start to link together. Therefore, the PPN films start to be formed as highly cross-linked hydrogen free polymers.[22,23] The peaks located at about 3480 cm-1 and 1600-1800 cm-1 region correspond to O-H and C=O stretch modes.[21,22,25,26] It is clearly observed from OES spectra that there are not any major peaks related to Oxygen spices. Therefore, these are probably due to post polymerization reactions of retained activated radicals with Oxygen or water vapor contents of atmosphere.[22,23]

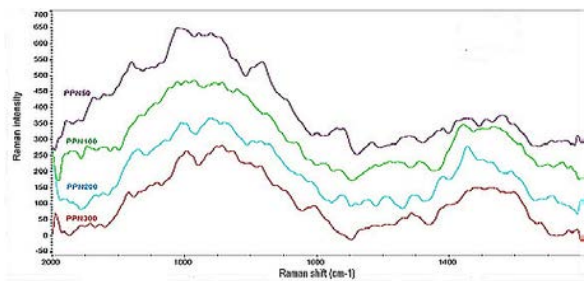

Figure 5. Raman spectra of the PPN films prepared in different powers.

Raman spectra of the PPN films (Figure 5) showed two broad peaks in 1300-1450 cm-1 and 1600-1850 cm-1 region due to the wide distribution of polymer chain lengths and less crystallinity.[27] Although, it is difficult to identify the bands because of overlapping, the former region is probably due to the structure disorder of polymer films.[28,29] In the later region, the peaks at around 1600 cm-1 are related to graphitic carbon and suggest the presence of sp2 bonds in the polymer structure.[22,23,27] Peaks at about 1630-1700 cm-1 and 1750-1800 cm-1 are attributed to carbonyl groups because of oxidation of the polymer film in contact with air.[29,30] By increasing the power, a slight shift to the lower wavenumbers was observed in 1600-1850 cm-1 region. It seems that this shift is related to a increasing of the conjugation length.[27,31]

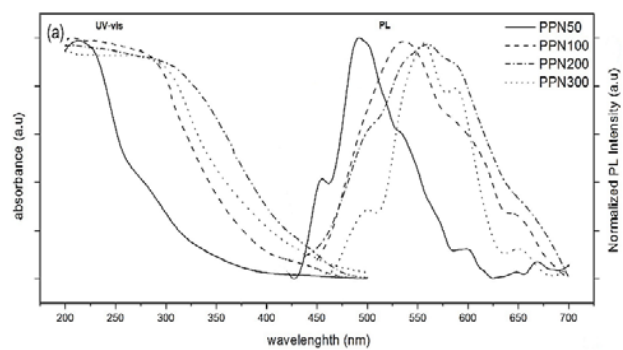

Figure 6a. Uv-vis absorption and PL spectra of PPN polymer films.

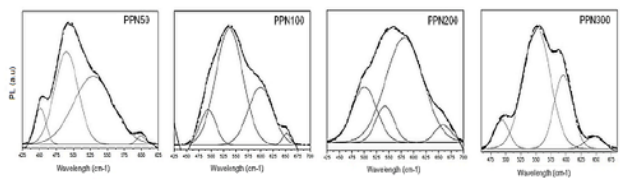

Figure 6b. PL spectra of PPN films fitted by Gaussian peaks.

UV-Vis and PL spectroscopies (Figure 6a) provided interesting information about the optical properties of the polymer films. UV-Vis spectrum of plasma polymerized naphthalene showed broad peaks at 225-325nm, which is attributed to the π-π* transition of the conjugated structure.[8,13,32] By increas-ing the power from 50 w to 200 w, absorption spectra showed a red-shift from 225 to 325 nm. It indicated that energy gap decreased as the plasma power increased. Increasing the plasma power caused a cross-linking increase and forming a longer conjugated sequence.[15] Further increasing the plasma power to 300 w caused a blue shift to about 300 nm that revealed an increase in the energy gap. It is attributed to the reduction of conjugation, probably due to double bond rupture of C=C because of energetic electrons bombardment at high powers. Moreover, PL spectra showed broad emission peaks in the visible region due to forming conjugated polymer chains with a wide distribution of lengths. For further analysis, the PL spectra were fitted by several Gaussian peaks (Figure 6b) corresponding to their major peaks and shoulders. The PL spectrum of the PPN50 included a peak located at 490 nm and three noticeable shoulders at 450, 525 and 600 nm. This profile was well fitted by four Gaussian peaks located at 450, 490, 528 and 600 nm. For PPN100, PL spectrum showed a new shoulder at 650 nm while the shoulder located at 450 nm vanished. The Gaussian profile indicated that the peak located at 490 nm decreased dramatically while the next peak was red shifted to 535nm and became dominated and the 600 nm peak increased as well. For PPN200 the trend of growing the longer emission peaks was followed and 600 and 650 nm emissions increased while 535 nm peak decreased. The PL spectrum of PPN300 showed a reverse behavior. Two longer wavelength shoulders decreased and the peak located at 550 nm increased.

The 450 nm emission might be due to the excited state of conjugated sequence. The 490 nm emission originated from Pi-stacks and peaks at longer wavelengths (at about 535, 600 and 650 nm) are due to excimeric emissions.[33,34] Emissions at 600 and 650 nm might be due to the inter-chain interaction between longer conjugated polymer chains. As plasma power and thus cross-linking increased, excimeric emissions became dominant. At higher powers, proper to cross-linking and polymers conjugation length, dominance of different excimeric emissions changed. For example, at 200w increasing the power caused dominance of longer conjugated chains thus the 600 nm emission became the major peak. On the other hand, at 300 w, because of energetic electron bombardment, conjugation decreased, therefore, the 530 nm peak became dominant again. Overall, PL spectra trend to shift to longer wavelengths by increasing the power due to increasing conjugation length except for 300 w. At 300w high energy electron bombardment prevents to form more conjugated polymer.

For fabrication of a light-emitting device, the plasma polymerized naphthalene was used as an emissive layer. The structure of the device mentioned previously. Figure 7 showed the EL spectra of PPN based devices. The turn on voltage of PPN devices was about 6-12 v.

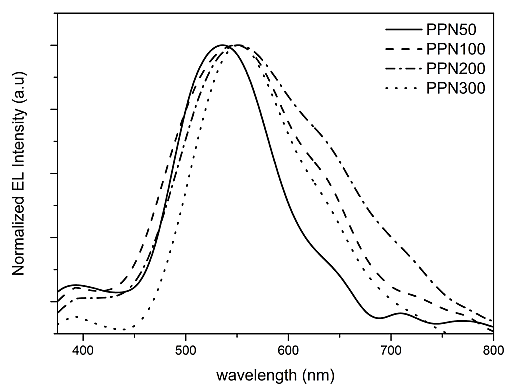

Figure 7. Normalized Electroluminescence spectra of the PPN based devices

The EL emission peaks were located at 535-550 nm. As the plasma power increased, a broadening of the emission bands occurred due to an increase in the complexity of the resulted polymers structure.[24] Like PL spectra, PPN300 showed a reverse behavior probably due to the fragmentation in the polymer structure because of high energy electron bombardment. In addition, the broadening caused a decrease in color purity.[35] All of the devices showed a stable spectra, however the device with the PPN50 emissive layer had more stable emission during the operation.

## Conclusion

In the present study, we used the PECVD method for the purpose of synthesizing an emissive layer. The Naphthalene monomer was used as starting material and introduced into the reaction chamber under a highly controlled condition. We also investigated the effect of the plasma power parameter on the films structure and optical properties. OES spectra showed C-H bond rupture and increasing Carbon atoms concentration as the plasma power increased. It shows increasing probability of forming more cross-linked polymer in higher powers that is in agreement with theory. The structure analysis revealed forming a conjugated polymer with a 3-D cross-linked structure. Increasing the power caused increasing both conjugation and cross-linking except for the 300 w power that the conjugation decreased. In addition, as the plasma power increased the optical properties showed two domains: below 200 w and higher than it. In the first domain the PL and UV-Vis spectra were red-shifted and became broader. These are the result of increasing the conjugation length and forming more complex polymers. Also, increasing the power and thus cross-linking caused rising the excimeric emissions at 535, 600 and 650 nm. In the second domain where the power increased to 300 w, high energy electron bombardment caused decreasing the conjugation and dominance of the lowest excimeric emission band at 535 nm. All of PPN films showed broad EL spectra centered at 535-550 nm. The emission spectra became broader with increasing the power due to forming a wide distribution of conjugation lengths in cross-linked chains. The EL spectra made a reverse behavior as the power increased to 300 w. The present study reported a new method for applying the plasma to form conjugated polymers and possibility of using the plasma power for engineering polymers with different structure and optical properties.